\documentclass[preprint,showpacs,preprintnumbers,amsmath,amssymb]{revtex4}

\usepackage{graphicx}
\usepackage{dcolumn}
\usepackage{bm}

\def\lesssim{\ \raise.3ex\hbox{$<$}\kern-0.8em\lower.7ex\hbox{$\sim$}\ }
\def\gesim{\ \raise.3ex\hbox{$>$}\kern-0.8em\lower.7ex\hbox{$\sim$}\ }
\begin{document}
\title{Evolution of Cooper pairs with zero-center-of-mass momentum and their first-order correlation function in a two-dimensional ultracold Fermi gas near the  observed Berezinskii-Kosterlitz-Thouless transition}
\author{Morio Matsumoto}\author{Daisuke Inotani}\author{Yoji Ohashi}
\affiliation{Faculty of Science and Technology, Keio University, 3-14-1 Hiyoshi, Kohoku-ku, Yokohama 223-8522, Japan.}
\begin{abstract}
We investigate the center-of-mass momentum distribution $n_{\bm Q}$ of Cooper pairs and their first-order correlation function $g_1(r)$ in a strongly interacting two-dimensional Fermi gas. Recently, the BKT (Berezinskii-Kosterlitz-Thouless) transition was reported in a two-dimensional $^6$Li Fermi gas, based on (1) the observations of anomalous enhancement of $n_{{\bm Q}={\bm 0}}$ [M. G. Ries, {\it et. al.}, Phys. Rev. Lett. {\bf 114}, 230401 (2015)], as well as (2) a power-law behavior of $g_1(r)$ [P. A. Murthy, {\it et. al.}, Phys. Rev. Lett. {\bf 115}, 010401 (2015)].  However, including pairing fluctuations within a $T$-matrix approximation (TMA), we show that these results can still be explained as strong-coupling properties of a {\it normal-state} two-dimensional Fermi gas. Our results indicate the importance of further experimental observations, to definitely confirm the realization of the BKT transition in this system. Since the BKT transition has been realized in a two-dimensional ultracold Bose gas, our results would be useful for the achievement of this quasi-long range order in an ultracold Fermi gas. 
\end{abstract}
\pacs{03.75.Ss, 03.75.-b, 67.85.Lm}
\maketitle
\section{Introduction}
\par
Since the achievement of the BKT (Berezinskii-Kosterlitz-Thouless) transition\cite{Berezinskii1,Berezinskii2,Kosterlitz1,Kosterlitz2} in a two-dimensional $^{87}$Rb Bose gas\cite{Dalibard1,Dalibard2,Schweikhard}, the possibility of this quasi-long range order\cite{Mermin,Hohenberg} in a Fermi gas has been explored both theoretically\cite{Botelho,Tempere,Klimin,Pietila,Watanabe,Bauer,Fischer,Matsumoto,Marsiglio,Matsumoto1,Chien,Salasnich,HuiHu} and experimentally\cite{Feld,Frohlich,Sommer,Murthy,Bloch}. Once the BKT transition is realized, using a tunable pairing interaction associated with a Feshbach resonance\cite{Timmermans,Chin}, we can examine physical properties of this two-dimensional Fermi superfluid, from the weak-coupling regime to the strong-coupling limit in a systematic manner. Since the BCS (Bardeen-Cooper-Schrieffer)-BEC (Bose Einstein condensation) crossover\cite{Eagles,Leggett,NSR,SadeMelo,Randeria2,OhashiGriffin,Perali,Gurarie,Giorgini,Ketterle} has been realized in the three dimensional case\cite{Regal,Zwierlein,Kinast,Bartenstein}, it is an interesting problem how this kind of crossover phenomenon occurs in the two-dimensional case\cite{Iskin}. 
\par
Recently, the realization of the BKT transition was reported in a strongly interacting two-dimensional $^6$Li Fermi gas\cite{Ries}. In this experiment, the center-of-mass momentum distribution $n_{\bm Q}$ of Cooper pairs was measured, by using a focusing technique\cite{Murthy} developed in an ultracold Bose gas\cite{Petrov,Druten,Tung}. The BKT transition temperature $T_{\rm BKT}^{\rm exp}$ was experimentally determined as the temperature below which the number $n_{{\bm Q}={\bm 0}}$ of Cooper pairs with zero center-of-mass momentum anomalously increases. Below $T_{\rm BKT}^{\rm exp}$, the power-law decay of the first-order correlation function of Cooper-pair bosons,
\begin{equation}
g_1(r)=\langle\Phi^\dagger({\bm r})\Phi(0)\rangle \sim {1 \over r^\eta},
\label{eq.1}
\end{equation}
was also reported as another evidence for the BKT transition\cite{Murthy1} (where $\Phi({\bm r})$ is the Bose field describing Cooper pairs). 
\par
However, the interpretation for these experiments\cite{Ries,Murthy1} still has room for discussion.  First, while the observed $T_{\rm BKT}^{\rm exp}$ increases with decreasing the strength of a pairing interaction, this tendency is opposite to the prediction by the BKT theory\cite{Botelho,Tempere,Klimin,Matsumoto}. Second, the observed exponent $\eta\simeq 1.4$ in Eq. (\ref{eq.1}) at $T_{\rm BKT}^{\rm exp}$ is much larger than the theoretical prediction, $\eta=0.25$\cite{Berezinskii1,Berezinskii2,Kosterlitz1,Kosterlitz2}. For the latter, this theoretical value has experimentally been observed in a two-dimensional $^{87}$Rb Bose gas\cite{Dalibard1}, as well as in an exciton-polariton condensate in a two-dimensional quantum well\cite{Yamamoto}. At present, the so-called universal jump of the superfluid density $\rho_{\rm s}$\cite{Nelson}, as well as vortex-antivortex pair annihilations\cite{Berezinskii1,Berezinskii2,Kosterlitz1,Kosterlitz2} (that are both characteristic of a BKT superfluid), have not been observed at $T_{\rm BKT}^{\rm exp}$. Thus, it is a crucial problem in the current stage of research whether the observed anomaly in $n_{{\bm Q}={\bm 0}}$, as well as the power-law behavior of $g_1(r)$ (apart from the value of the exponent), are enough to conclude the BKT phase transition, or further evidence is necessary in order to unambiguously confirm the achievement of this two-dimensional Fermi superfluid.
\par
In this paper, including pairing fluctuations within the framework of a $T$-matrix approximation (TMA)\cite{Tsuchiya,Perali}, we investigate strong-coupling corrections to the number $n_{\bm Q}$ of Cooper pairs with center-of-mass momentum ${\bm Q}$, as well as their first-order correlation function $g_1(r)$, in a normal-state two-dimensional ultracold Fermi gas. In the three-dimensional case, TMA has extensively been used to examine the so-called pseudogap phenomenon associated with strong pairing fluctuations in the BCS-BEC crossover region\cite{Tsuchiya,Levin2009,Tsuchiya2010,HuiHu2010,Gaebler2010}. TMA has also been applied to a two-dimensional $^{40}$K normal Fermi gas in a harmonic trap\cite{Watanabe}, to successfully explain the pseudogap size observed by a photoemission-type experiment\cite{Feld}. Thus, although TMA cannot deal with the BKT transition\cite{Mermin,Hohenberg,Pietila}, we can expect that it is still useful for the study of strong-coupling physics in the {\it normal state} of a two-dimensional Fermi gas. 
\par
Within the framework of TMA, we examine whether the anomalous amplification of $n_{{\bm Q}={\bm 0}}$ observed in a two-dimensional $^6$Li Fermi gas\cite{Ries} is unique to the BKT phase transition, or it can be understood as a strong-coupling phenomenon in the normal state. In a previous paper\cite{Matsumoto1}, we have briefly discussed this problem at an intermediate coupling strength. In this paper, we extend this to a wide region with respect to the interaction strength. In addition to this, we also evaluate $g_1(r)$ within the same TMA, to see whether or not the BKT transition is necessary to explain the observed large value of the exponent, $\eta\simeq 1.4$. 
\par
We briefly note that, Ref. \cite{Chien} recently discussed a two-dimensional Fermi gas on the viewpoint of quasi-condensate, within a static approximation to the amplitude of pairing fluctuations. In this paper, we fully deal with pairing fluctuations within TMA, without employing any further approximation to them.
\par
This paper is organized as follows. In Sec. II, we explain a strong-coupling $T$-matrix approximation (TMA) in a two-dimensional normal Fermi gas. In Sec. III, we examine $n_{\bm{Q}={\bm 0}}$, to see how the temperature around which $n_{\bm{Q}={\bm 0}}$ starts to remarkably increase is close to $T_{\rm BKT}^{\rm exp}$ observed in a two-dimensional $^6$Li Fermi gas\cite{Ries}. In Sec. IV, we consider $g_1(r)$. Following the experimental prescription\cite{Murthy1}, we extract the exponent $\eta$ in Eq. (\ref{eq.1}) from the calculated $g_1(r)$, to examine whether or not the observed value $\eta\simeq 1.4$\cite{Murthy1} can be explained in the normal state, without assuming the BKT transition. Throughout this paper, we take $\hbar =k_{\rm B}=1$, and the system area is taken to be unity, for simplicity.
\par
\section{Formulation}
\par
We consider a two-dimensional uniform Fermi gas, described by the BCS Hamiltonian, 
\begin{align}
H=\sum_{\bm{p},\sigma}\xi_{\bm{p}}
c^{\dagger}_{\bm{p},\sigma}c_{\bm{p},\sigma}
-U
\sum_{\bm{p},\bm{p}',\bm{Q}}
c^{\dagger}_{\bm{p}+\bm{Q}/2,\uparrow}
c^{\dagger}_{-\bm{p}+\bm{Q}/2,\downarrow}
c_{-\bm{p}'+\bm{Q}/2,\downarrow}
c_{\bm{p}'+\bm{Q}/2,\uparrow}.
\label{eq2.1}
\end{align}
Here, $c^{\dagger}_{\bm{p},\sigma}$ is the creation operator of a Fermi atom with pseudospin $\sigma =\uparrow, \downarrow$, and the two-dimensional momentum ${\bm p}=(p_x,p_y)$. $\xi_{\bm p}=p^2/(2m)-\mu$ is the kinetic energy of a Fermi atom, measured from the Fermi chemical potential $\mu$, where $m$ is an atomic mass. $-U$ $(<0)$ is a tunable pairing interaction associated with a Feshbach resonance\cite{Timmermans,Chin}. As usual, we conveniently measure the interaction strength in terms of the the $s$-wave scattering length $a_{\rm 2D}$, which is related to $U$ as, in the two-dimensional case\cite{Watanabe,Morgan}, 
\begin{equation}
\frac{1}{U}=\frac{m}{2\pi}\ln
\left(
k_{\rm F}a_{\rm 2D}
\right)
+\sum_{p\ge k_{\rm F}}\frac{m}{p^2}, 
\label{eq2.2}
\end{equation}
where $k_{\rm F}$ is the Fermi momentum. In this scale, the weak coupling side and the strong-coupling side are characterized as $\ln(k_{\rm F}a_{\rm 2D}) \gtrsim 0$, and $\ln(k_{\rm F}a_{\rm 2D}) \lesssim 0$, respectively.
\par
\begin{figure}[t]
\begin{center}
\includegraphics[keepaspectratio,scale=0.7]{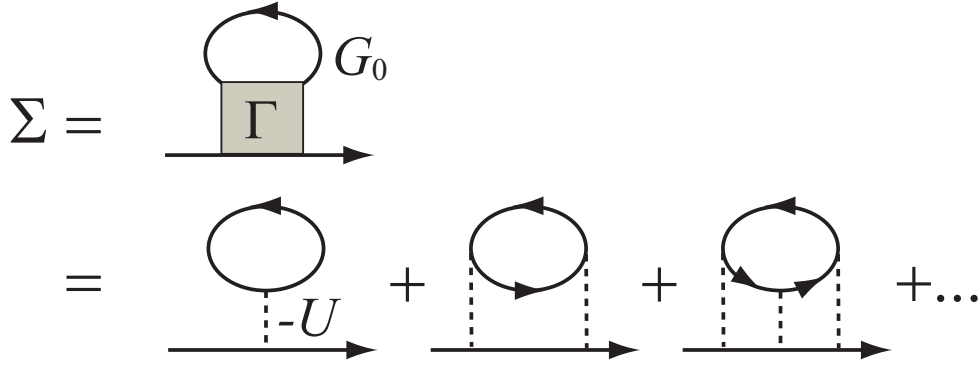}
\caption{Self-energy correction $\Sigma({\bm p},i\omega_n)$ in a $T$-matrix approximation (TMA). The solid line is the single-particle Green's function $G_0$ for a free Fermi gas. $-U$ is a pairing interaction. $\Gamma$ is the TMA particle-particle scattering matrix.}
\label{fig1}       
\end{center}
\end{figure}
\par
Strong-coupling effects on Fermi single-particle excitations are well described by the self-energy correction $\Sigma({\bm p},i\omega_n)$ in the single-particle thermal Green's function, given by
\begin{equation}
 G(\bm{p},i\omega_{n})=\frac{1}{i\omega_{n}-\xi_{\bm{p}}-\Sigma(\bm{p},i\omega_{n})}, \label{eq2.3}
\end{equation}
where $\omega_{n}$ is the fermion Matsubara frequency. In TMA, the self-energy is diagrammatically described as Fig. \ref{fig1}, which gives\cite{Perali,Tsuchiya}
\begin{equation}
\Sigma({\bm p},i\omega_n)=
T\sum_{\bm{Q},i\nu_{n}}
\Gamma(\bm{Q},i\nu_{n})G_{0}(\bm{Q}-\bm{p},i\nu_{n}-i\omega_{n}).
\label{eq2.4}
\end{equation}
Here, $\nu_{n}$ is the boson Matsubara frequency, and
\begin{equation}
G_0(\bm{p},i\omega_{n})=
{1 \over
i\omega_{n}-\xi_{\bm{p}}
}
\label{eq2.4b}
\end{equation}
is the single-particle Green's function for a free Fermi gas. In Eq. (\ref{eq2.4}),
\begin{equation}
\Gamma(\bm{Q},i\nu_{n})=-\frac{U}{1-U\Pi(\bm{Q},i\nu_{n})}
\label{eq2.5}
\end{equation}
is the TMA particle-particle scattering matrix, where 
\begin{eqnarray}
\Pi(\bm{Q},i\nu_{n})
&=&
T\sum_{\bm{p},i\omega_{n}}G_{0}(\bm{p}+{\bm{Q}}/2,i\nu_{n}+i\omega_{n})G_{0}(-\bm{p}+{\bm{Q}}/2,-i\omega_{n})
\nonumber
\\
&=&
\sum_{\bm p}
{1-f(\xi_{{\bm p}+{\bm Q}/2})-f(\xi_{-{\bm p}+{\bm Q}/2})
\over
\xi_{{\bm p}+{\bm Q}/2}+\xi_{-{\bm p}+{\bm Q}/2}-i\nu_n
}, 
\label{eq2.6}
\end{eqnarray}
is the lowest-order pair-correlation function, describing two-dimensional pairing fluctuations, where $f(x)$ is the Fermi distribution function. 
\par
The Fermi chemical potential $\mu$ is determined from the equation for the total number $N$ of Fermi atoms, given by
\begin{equation}
N=2T\sum_{\bm{p},i\omega_{n}}G(\bm{p},i\omega_{n}),
\label{eq2.7}
\end{equation}
where the TMA Green's function $G({\bm p},i\omega_n)$ in Eq. (\ref{eq2.3}) is used. 
\par
\begin{figure}[t]
\begin{center}
\includegraphics[keepaspectratio,scale=0.5]{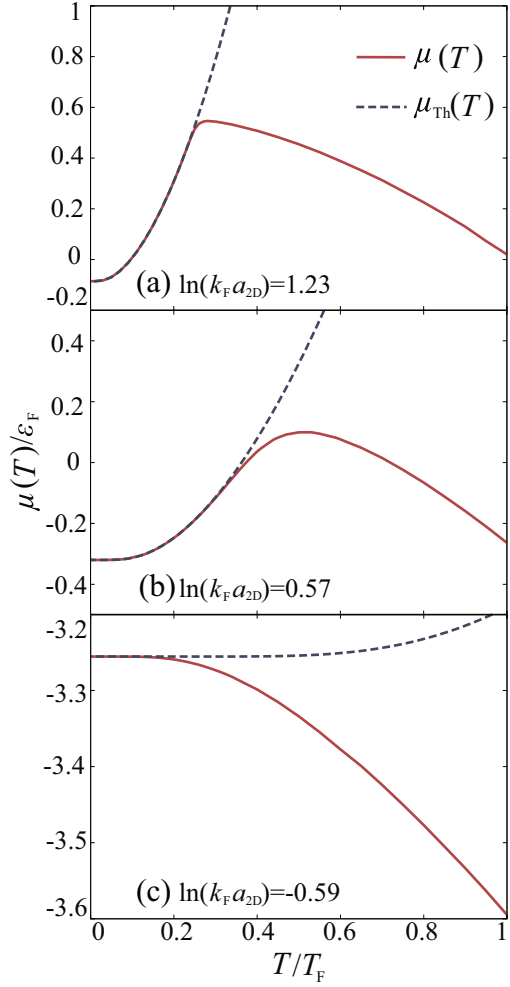}
\caption{(Color online) Calculated Fermi chemical potential $\mu$ in a two-dimensional Fermi gas, as a function of temperature. $\mu_{\rm Th}$ is the chemical potential which satisfies the Thouless criterion in Eq. (\ref{eq2.7b}). $T_{\rm F}$ and $\varepsilon_{\rm F}$ are the Fermi temperature and Fermi energy, respectively. We note that $\mu(T)$ is always smaller than $\mu_{\rm Th}(T)$ when $T>0$.}
\label{fig2}       
\end{center}
\end{figure}

To simply see that TMA does not give a finite superfluid phase transition temperature $T_{\rm c}$ in the two-dimensional case, we note that the equation for $T_{\rm c}$ in a three-dimensional Fermi superfluid is conveniently obtained from the Thouless criterion\cite{Thouless}, stating that the superfluid instability occurs when the particle-particle scattering matrix $\Gamma({\bm Q},i\nu_n)$ in Eq. (\ref{eq2.5}) has a pole at ${\bm Q}=\nu_n=0$, as
\begin{equation}
1-U\Pi(0,0)=0.
\label{eq2.7b}
\end{equation}
In the BCS-BEC crossover region, one solves the $T_{\rm c}$-equation (\ref{eq2.7b}), together with the number equation (\ref{eq2.7}), to self-consistently determine $T_{\rm c}$ and $\mu$. At $T_{\rm c}$, noting that $\Gamma({\bm Q},i\nu_n=0)$ around ${\bm Q}={\bm 0}$ has the form
\begin{equation}
\Gamma({\bm Q},0)\simeq 
2\left(
{\partial^2 \Pi({\bm Q},0) \over \partial Q^2}
\right)^{-1}_{{\bm Q}={\bm 0}}
{1 \over Q^2}\equiv{\gamma \over Q^2},
\label{eq2.7c}
\end{equation}
one can approximate the TMA self-energy in Eq. (\ref{eq2.4}) to
\begin{eqnarray}
\Sigma({\bm p},i\omega_n)
&=&
T\sum_{\bm{Q}}\Gamma(\bm{Q},0)G_{0}(\bm{Q}-\bm{p},-i\omega_{n})
+
T\sum_{\bm{Q},\nu_n\ne 0}\Gamma(\bm{Q},i\nu_n)G_{0}(\bm{Q}-\bm{p},i\nu_n-i\omega_{n})
\nonumber
\\
&\simeq&
T\gamma G_0(-{\bm p},-i\omega_n)\sum_{\bm Q}{1 \over Q^2}
+
T\sum_{\bm{Q},\nu_n\ne 0}\Gamma(\bm{Q},i\nu_n)G_{0}(\bm{Q}-\bm{p},i\nu_n-i\omega_{n}).
\label{eq2.7d}
\end{eqnarray}
While Eq. (\ref{eq2.7d}) converges in the three-dimensional case, the summation $\sum_{\bm Q}(1/Q^2)$ logarithmically diverges in two dimension, the latter of which means that the coupled equations (\ref{eq2.7}) with (\ref{eq2.7b}) are never satisfied simultaneously. As a result, as shown in Fig. \ref{fig2}, the Fermi chemical potential $\mu$ which is determined from the number equation (\ref{eq2.7}) is always smaller than the chemical potential $\mu_{\rm Th}$ which satisfies the Thouless criterion in Eq. (\ref{eq2.7b})\cite{Tokumitu,Schmitt}. The essence of this vanishing $T_{\rm c}$ is just the same as the Hohenberg's theorem\cite{Hohenberg}, stating that the superfluid long-range order in the two-dimensional case is completely destroyed by low-energy superfluid fluctuations. 
\par
\par
\begin{figure}[t]
\begin{center}
  \includegraphics[keepaspectratio,scale=0.5]{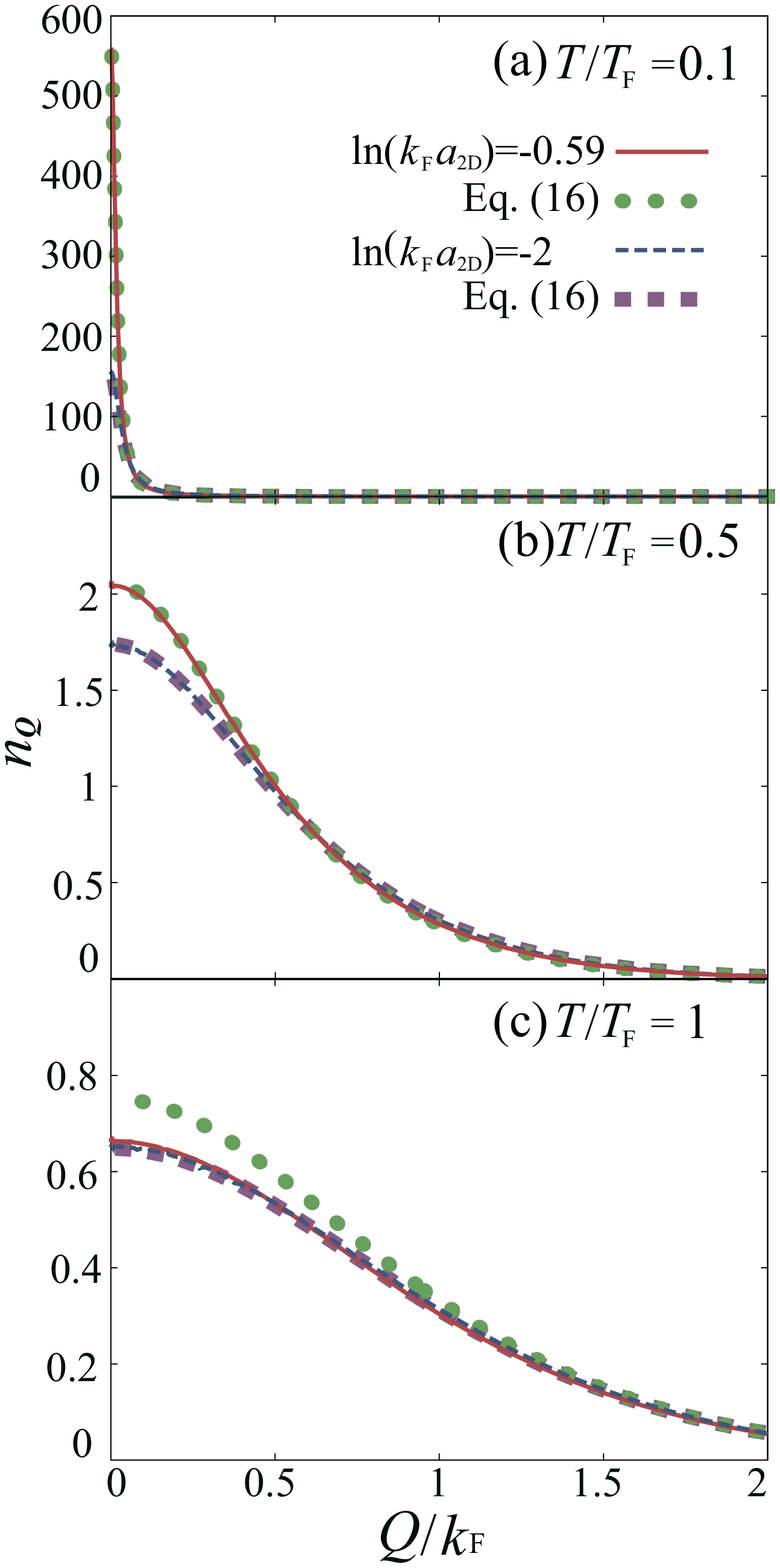}
\caption{(Color online) Calculated center-of-mass momentum distribution function $n_{\bm Q}$ of Cooper pairs in Eq. (\ref{eq2.11}) in the strong-coupling side ($\ln(k_{\rm F}a_{\rm 2D})<0$). The filled circles (squares) show Eq. (\ref{eq3.1a}) when $\ln(k_{\rm F}a_{\rm 2D})=-0.59$ ($\ln(k_{\rm F}a_{\rm 2D})=-2$).}
\label{fig3}       
\end{center}
\end{figure}
\par
To introduce the center-of-mass momentum distribution $n_{\bm Q}$ of Cooper pairs in the present TMA formalism, we conveniently divide the number equation (\ref{eq2.7}) into the sum of the number $N_0$ of a free Fermi atoms and the fluctuation correction $\delta N$, as $N=N_0+\delta N$. Each component is given by,
\begin{equation}
N_0=2T\sum_{\bm{p},i\omega_{n}}G_0(\bm{p},i\omega_{n})
=2\sum_{\bm p}f(\xi_{\bm p}),
\label{eq2.8}
\end{equation}
\begin{equation}
\delta N=2T\sum_{\bm{p},i\omega_{n}}
\left[G(\bm{p},i\omega_{n})-G_{0}(\bm{p},i\omega_{n})\right].
\label{eq2.9}
\end{equation}
Equation (\ref{eq2.9}) involves the number of stable Cooper pairs, as well as the contribution of the so-called scattering states\cite{NSR}, the latter of which physically describes effects of fluctuations in Cooper channel. Although the latter is somehow different from the ``number" of particles because it is known to become negative in the three-dimensional case\cite{NSR}, this paper simply deals with scattering states as {\it fluctuating} Cooper-pair bosons. In this case, when we relate $n_{\bm Q}$ to $\delta N$ as $\delta N=2\sum_{\bm Q}n_{\bm Q}$ (Note that one boson counts as two Fermi atoms.), we obtain from Eq (\ref{eq2.9}),
\begin{eqnarray}
n_{\bm{Q}}=T\sum_{i\nu_{n}}
\Gamma(\bm{Q},i\nu_{n}) T\sum_{\bm{p}, i\omega_{n}}
G_{0}(\bm{Q}-\bm{p},i\nu_{n}-i\omega_{n})
G_{0}(\bm{p},i\omega_{n})
G(\bm{p},i\omega_{n}).
\label{eq2.11}
\end{eqnarray}
Indeed, Eq. (\ref{eq2.11}) has the bosonic character in the strong-coupling side (where $\mu<0$), as
\begin{equation}
n_{\bm{Q}}
\simeq
Z n_{\rm B}\left(\frac{Q^2}{2M}-\mu_{\rm B}\right)~~(T\ll|\mu|),
\label{eq3.1a} 
\end{equation}
where $n_{\rm B}(x)$ is the Bose distribution function. (We summarize the derivation of Eq. (\ref{eq3.1a}) in the Appendix.) Equation (\ref{eq3.1a}) just equals the momentum distribution of a Bose gas with the molecular mass, $M=2m$, the Bose chemical potential potential, 
\begin{equation}
\mu_{\rm B} = -2|\mu|
\ln{ \left(\frac{2|\mu|} {E^{\rm 2D}_{\rm bind}} \right) }~~(<0)
\label{eq2.12}
\end{equation}
(where $E^{\rm 2D}_{\rm bind}=1/(ma^2_{\rm 2D})$ is the binding energy of a two-body bound state\cite{Randeria}), and the renormalization factor,
\begin{equation}
Z={|\mu| \over |\mu|+\varepsilon_{\rm F}},
\label{eq2.12b}
\end{equation} 
which effectively describes deviation from an ideal Bose gas due to a finite value of the pairing interaction. In the strong-coupling limit (where the Fermi chemical potential $\mu$ approaches half the binding energy as $\mu=-E^{\rm 2D}_{\rm bind}/2\to -\infty$), the renormalization factor $Z$ in Eq. (\ref{eq2.12b}) is reduced to unity, so that Eq. (\ref{eq3.1a}) simply gives the ordinary momentum distribution of an ideal Bose gas, as expected.
\par
As shown in Fig. \ref{fig3}, $n_{\bm Q}$ in Eq. (\ref{eq2.11}) is well described by Eq. (\ref{eq3.1a}), not only deep inside the strong-coupling regime ($\ln(k_{\rm F}a_{\rm 2D})=-2$), but also in the region relatively close to the intermediate coupling region ($\ln(k_{\rm F}a_{\rm 2D})=-0.59$) when $T\ll T_{\rm F}$ (where $T_{\rm F}$ is the Fermi temperature). This indicates that the present definition of $n_{\bm Q}$ really has the meaning of the momentum distribution of Cooper-pair ``bosons" in the strong-coupling side $\ln(k_{\rm F}a_{\rm 2D})\lesssim0$. 
\par
Once $n_{\bm Q}$ is determined, the first-order correlation function $g_1(r)$ of Cooper-pair bosons is immediately obtained from the Fourier transformation of $n_{\bm Q}$ in Eq. (\ref{eq2.11}) as
\begin{equation}
g_{1}(r)=\sum_{\bm Q} n_{\bm{Q}}e^{i\bm{Q}\cdot \bm{r}}.
\label{eq2.13}
\end{equation}
\par
Before ending this section, we discuss the validity of the TMA in the two-dimensional case. When $\mu\simeq \mu_{\rm Th}$ (see Fig. \ref{fig2}), pairing fluctuations described by the TMA particle-particle scattering matrix $\Gamma({\bm Q},i\nu_n)$ in Eq. (\ref{eq2.5}) are strongly enhanced around ${\bm Q}=\nu_n=0$, leading to a pseudogap in the single-particle density of states $\rho(\omega)$\cite{Matsumoto,Marsiglio}. Indeed, in such situation, one may approximate the self-energy in Eq. (\ref{eq2.4}) to\cite{Tsuchiya,Levin2009}
\begin{equation}
\Sigma({\bm p},i\omega_n)\simeq
G_{0}(-\bm{p},-i\omega_{n})
T\sum_{{\bm Q}, i\nu_n} 
\Gamma \left( {\bm Q}, i\nu_n \right).
\label{eq2.13a}
\end{equation}
The TMA single-particle Green's function in Eq. (\ref{eq2.3}) then has the same form as the diagonal component of the BCS Green's function\cite{Schrieffer} as,
\begin{equation}
G(\bm{p},i\omega_{n})=
-\frac{i\omega_{n}+\xi_{\bm{p}}}{\omega_{n}^2+\xi_{\bm{p}}^2+\Delta_{\rm PG}^2}.\label{eq2.13b}
\end{equation}
As in the ordinary BCS theory, Eq. (\ref{eq2.13b}) gives a pseudogap in $\rho(\omega)$, with the gap size,
\begin{equation}
\Delta_{\rm PG}=\sqrt{-T\sum_{{\bm Q}, i\nu_n}\Gamma \left( {\bm Q}, i\nu_n \right)},
\label{eq2.13c}
\end{equation}
when $\mu>0$. Since the ${\bm Q}$-summation diverges in the two-dimensional case when $\mu=\mu_{\rm Th}$ is satisfied (see the discussion below Eq. (\ref{eq2.7d})), the pseudogap would become remarkable in the temperature region where $\mu\simeq\mu_{\rm Th}$ in Fig. \ref{fig2}.
\par
However, such a strong-coupling phenomenon (that is further enhanced by the two-dimensionality of the system) is completely ignored in the TMA particle-particle scattering matrix $\Gamma({\bm Q},i\nu_n)$ in Eq. (\ref{eq2.5}), because the bare Green's function $G_0$ is still used there. Since an energy gap should suppress pairing fluctuations, TMA is considered to overestimate strong-coupling effects in the low temperature region of the weak-coupling regime where $\mu\simeq\mu_{\rm Th}$. This would also affect the center-of-mass momentum distribution function $n_{\bm Q}$ of Cooper pairs through the factor $\Gamma({\bm Q},i\nu_n)$ appearing in Eq. (\ref{eq2.11}).
\par
When $\mu<0$ in the strong-coupling side ($\ln(k_{\rm F}a_{\rm 2D})<0$), the approximate Green's function in Eq. (\ref{eq2.13a}) has the pseudogap with the gap size being equal to, not $\Delta_{\rm PG}$, but $E_{\rm PG}=\sqrt{|\mu|^2+\Delta_{\rm PG}^2}$. In this regime, this effect is partially taken into account in the TMA particle-particle scattering matrix $\Gamma({\bm Q},i\nu_n)$, because the bare Green's function $G_0$ in Eq. (\ref{eq2.4b}) has a finite energy gap $|\mu|$ when $\mu<0$. In the strong-coupling limit, because $|\mu|\gg\Delta_{\rm PG}$, one finds $E_{\rm PG}\to |\mu|$. 
\par
Thus, in this paper, we restrict our TMA analyses to the strong-coupling side, $\ln(k_{\rm F}a_{\rm 2D})<0$, where $\mu<0$. As mentioned previously, $n_{\bm Q}$ also has the required bosonic character in this regime. To extend our analyses to the weak-coupling side, we need to include strong-coupling corrections to the particle-particle scattering matrix $\Gamma({\bm Q},i\nu_n)$ beyond TMA, such as the self-consistent $T$-matrix approximation\cite{Bauer} (where the dressed Green's function is used in $\Gamma({\bm Q},i\nu_n)$). This extension remains as our future problem.
\par
\begin{figure}
\includegraphics[keepaspectratio,scale=0.5]{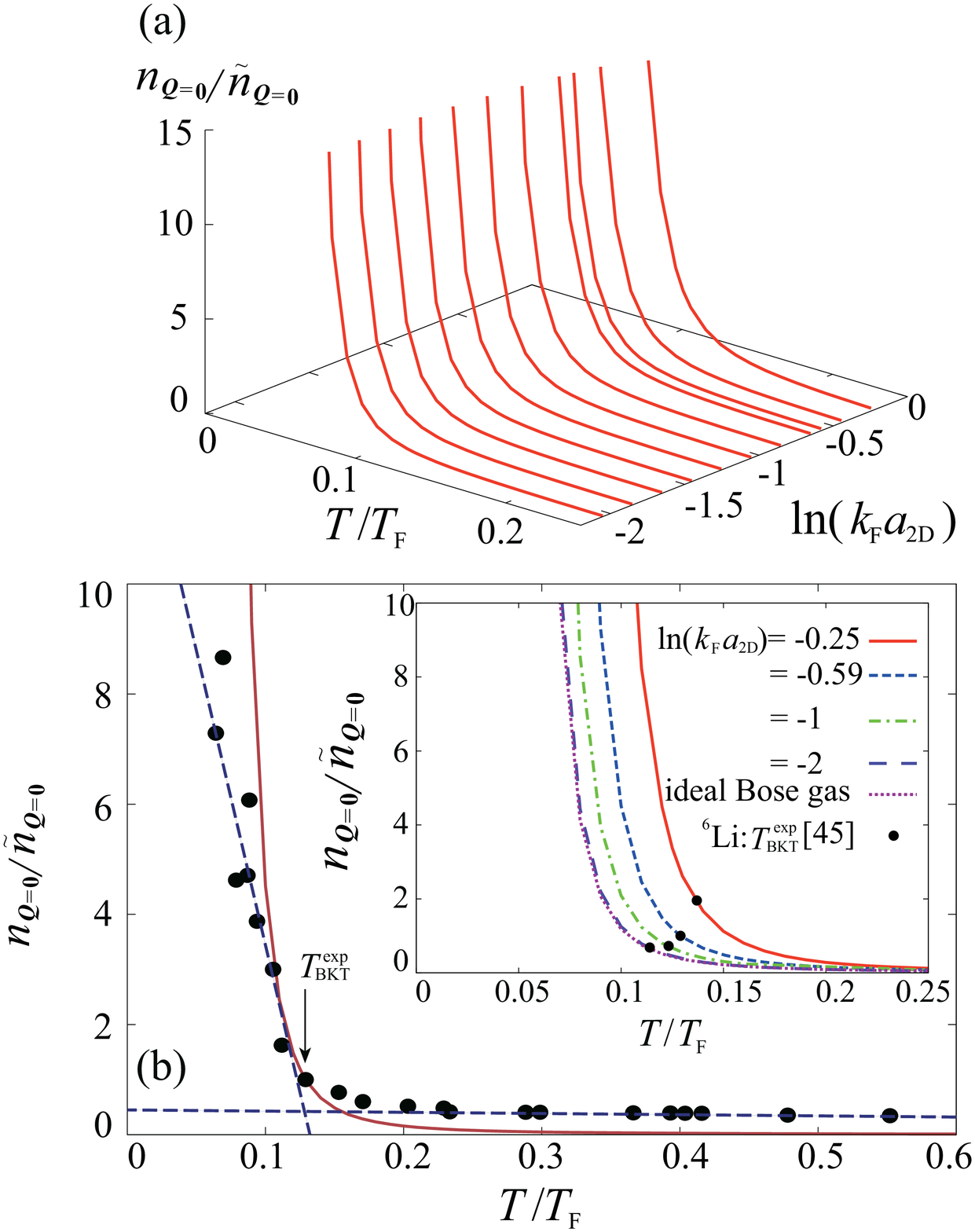}
\caption{(Color online) (a) Calculated number $n_{{\bm Q }={\bm 0}}$ of Cooper pairs with zero center-of-mass momentum (${\bm Q}={\bm 0}$) in TMA. In this figure, $n_{{\bm Q }={\bm 0}}$ is normalized by the value ($\equiv {\tilde n}_{{\bm Q }={\bm 0}}$) at the observed BKT phase transition temperature $T_{\rm BKT}^{\rm exp}=0.129T_{\rm F}$ when $\ln{(k_{\rm F}a_{\rm 2D})}=-0.59$\cite{Ries}. (b) Comparison of the calculated $n_{{\bm Q }={\bm 0}}$ (solid line) with the experimental result on a $^6$Li Fermi gas at $\ln{(k_{\rm F}a_{\rm 2D})}=-0.59$ (filled circles)\cite{Ries}. In this experiment, low temperature data and high temperature data are approximated to two $T$-linear functions (dashed lines in Fig. \ref{fig4}(b)), to determine $T_{\rm BKT}^{\rm exp}$ as the temperature at their intersection. The inset shows the temperature dependence of $n_{{\bm Q }={\bm 0}}$ in TMA around $T_{\rm BKT}^{\rm exp}$ (filled circles)\cite{Ries}. In this inset ``ideal Bose gas" is the momentum distribution function of a two-dimensional ideal gas of $N/2$ molecular bosons with the molecular mass $M=2m$.
}
\label{fig4}
\end{figure}

\section{The number $n_{{\bm Q}={\bm 0}}$ of Cooper pairs with zero center of mass momentum}
\par
Figure \ref{fig4}(a) shows the number $n_{{\bm Q}={\bm 0}}$ of Cooper pairs with zero center-of-mass momentum. In this figure, we see that $n_{{\bm Q}={\bm 0}}$ is anomalously enhanced at low temperatures, which is in good agreement with the recent experiment on a two-dimensional $^6$Li Fermi gas\cite{Ries}, as shown in Fig. \ref{fig4}(b). In this experiment\cite{Ries}, the BKT phase transition temperature $T_{\rm BKT}^{\rm exp}$ is determined as the temperature below which a large number of Cooper-pair bosons start to occupy the zero center-of-mass-momentum state. For this purpose, this experiment first fits the low temperature data ($T/T_{\rm F}\lesssim 0.1$) and high temperature data ($T/T_{\rm F}\gg 0.1$) to two linear functions as shown in Fig. \ref{fig4}(b) (dashed line), to determine $T_{\rm BKT}^{\rm exp}$ as the temperature at their intersection. Although this experimental prescription seems difficult to directly apply to the calculated $n_{{\bm Q}={\bm 0}}$ (because it actually does not exhibit linear temperature dependence in the low temperature region, as well as in the high temperature region), Fig. \ref{fig4}(b) clearly shows that our result also starts to remarkably increase with decreasing the temperature when $T\lesssim T_{\rm BKT}^{\rm exp}$. As shown in the inset in Fig. \ref{fig4}(b), this agreement is also obtained for other coupling strengths in the strong-coupling side.
\par
\begin{figure}
\includegraphics[keepaspectratio,scale=0.3]{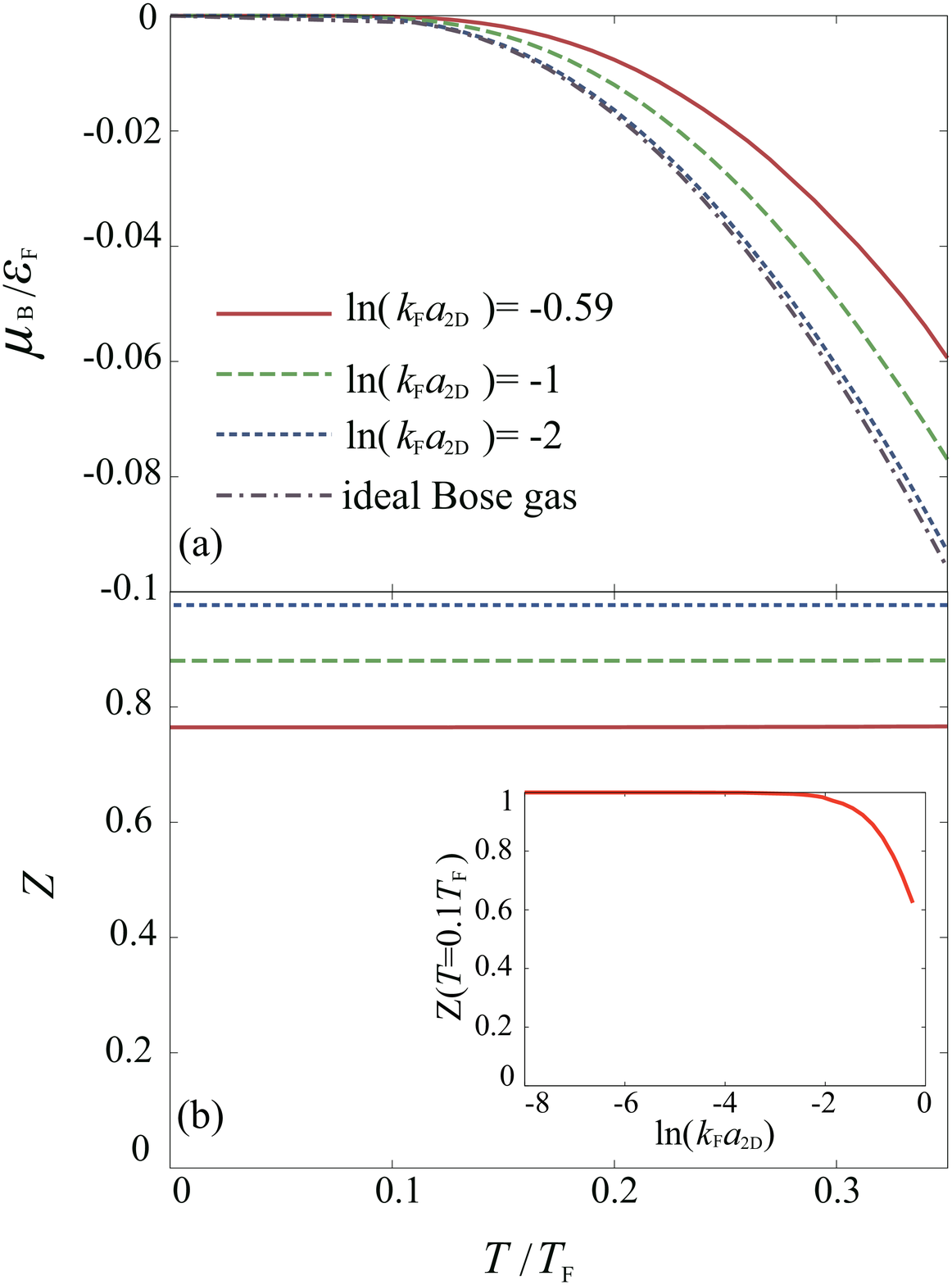}
\caption{(Color online) (a) Bose chemical potential $\mu_{\rm B}$ in Eq. (\ref{eq2.12}). In this figure, ``ideal Bose gas" shows the chemical potential in a two-dimensional ideal Bose gas, consisting of $N/2$ molecules with the molecular mass $M=2m$. (b) Renormalization factor $Z$ in Eq. (\ref{eq2.12b}), as a function of temperature. The inset shows $Z$ at $T=0.1T_{\rm F}$.}
\label{fig5}
\end{figure}  
\par
However, we emphasize that the remarkable increase of $n_{{\bm Q}={\bm 0}}$ shown in Fig. \ref{fig4}(a) is nothing to do with the BKT instability, because TMA does not give any superfluid phase transition in the two-dimensional case. Instead, as shown in the inset in Fig. \ref{fig4}(b), the low temperature behavior of the TMA $n_{{\bm Q}={\bm 0}}$ at $\ln(k_{\rm F}a_{\rm 2D})=-2$ is well described by the momentum distribution function in a two-dimensional ideal Bose gas (``ideal Bose gas" in the inset in Fig. \ref{fig4}(b)), consisting of $N/2$ molecular bosons with a molecular mass $M=2m$. In a two-dimensional ideal Bose gas, while the Bose-Einstein condensation is forbidden at finite temperatures, the Bose chemical potential $\mu_{\rm B}$ becomes very small at low temperatures (see Fig. \ref{fig5}(a)), which naturally causes the remarkable amplification of the number $n_{{\bm Q}={\bm 0}}^{\rm ideal}$ of ideal bosons at ${\bm Q}={\bm 0}$ as
\begin{equation}
n^{\rm ideal}_{{\bm Q}={\bm 0}}=
{\displaystyle 1 \over e^{|\mu_{\rm B}|/T}-1}\simeq 
{T \over |\mu_{\rm B}|}\gg 1.
\label{eq2.13d}
\end{equation}
\par
To explain the interaction dependence of $T_{\rm BKT}^{\rm exp}$ seen in the inset in Fig. \ref{fig4}(b) within the framework of TMA, we recall that the low temperature behavior of $n_{{\bm Q}={\bm 0}}$ in the strong-coupling side is well described by Eq. (\ref{eq3.1a}) (see Fig. \ref{fig3}). Using this, we can rewrite the number equation (\ref{eq2.7}) as
\begin{equation}
{N \over 2}=Z\sum_{\bm Q}n_{\rm B}
\left({Q^2 \over 2M}-\mu_{\rm B}\right)
=
\int_0^\infty {QdQ \over 2\pi}n_{\rm B}
\left({Q^2 \over 2(ZM)}-\mu_{\rm B}\right),
\label{eq2.13e}
\end{equation}
where, for simplicity, we have ignored the contribution $N_0$ of free Fermi atoms, by using the fact that $\mu<0$ in the strong-coupling side. Equation (\ref{eq2.13e}) is just the number equation in an ideal Bose gas, consisting of $N/2$ bosons with an {\it effective} molecular mass $M^*=ZM$. As shown in Fig. \ref{fig5}(b), the renormalization factor $Z$ becomes small with decreasing the interaction strength, reflecting that {\it fluctuations} of Cooper-pair bosons gradually become important. Thus, the effective mass $M^*=ZM$ becomes smaller for a weaker pairing interaction. Then, since the Bose chemical potential potential $\mu_{\rm B}$ becomes close to zero from higher temperatures in the case of lighter boson mass $M^*$ (see Figs. \ref{fig5}(a) and \ref{fig5}(b)), the remarkable enhancement of $n_{{\bm Q}={\bm 0}}$ starts to occur from higher temperatures for a weaker pairing interaction. The inset in Fig. \ref{fig4}(b) indicates that this interaction dependence of $n_{{\bm Q}={\bm 0}}$ well explains the behavior of $T_{\rm BKT}^{\rm exp}$ observed in a $^6$Li Fermi gas\cite{Ries}. 
\par
We briefly note that the interaction dependence of the observed $T_{\rm BKT}^{\rm exp}$ is quite opposite to the prediction by the BKT theory\cite{Botelho,Tempere,Klimin,Matsumoto}, where the BKT phase transition temperature is lowered, as one passes through the intermediate coupling regime from the strong-coupling side. At this stage, it is unclear whether the observed behavior of $T_{\rm BKT}^{\rm exp}$ can be explained by improving the current BKT theory or not, our results indicate that it can be understood as a normal-state property of a strongly interacting two-dimensional Fermi gas, at least in the strong-coupling side. 
\par
\begin{figure}
\includegraphics[keepaspectratio,scale=0.5]{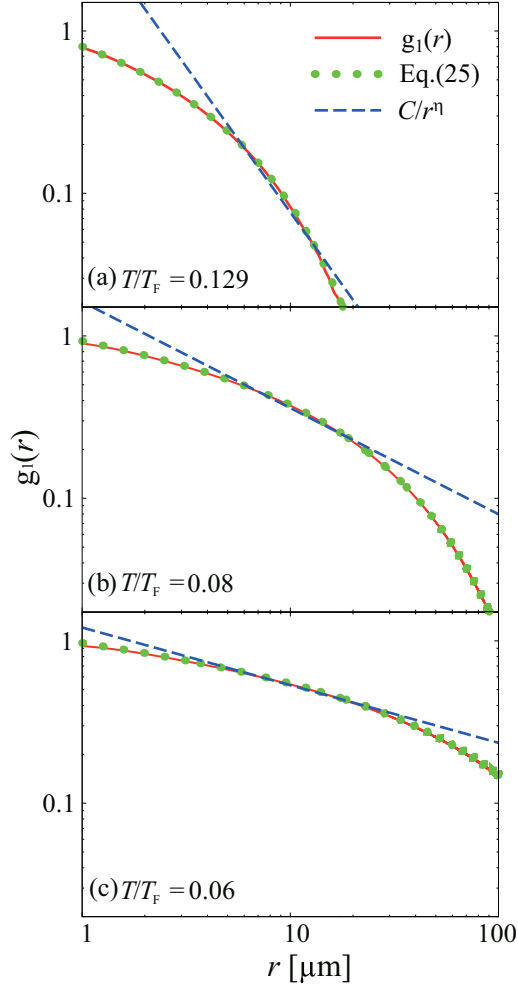}
\caption{(Color online) Calculated TMA first-order correlation function $g_{1}(r)$ of Cooper-pair bosons. We take $\ln{(k_{\rm F}a_{\rm 2D})}=-0.59$. The dashed line is the fitting result by the least squares method, when the fitting function $g_1(r)=C/r^\eta$ is assumed in the spatial region, $5{\mu{\rm m}}\le r\le 25{\mu{\rm m}}$. (In this region, Ref. \cite{Murthy1} experimentally evaluates $\eta$.) The filled circles show Eq. (\ref{eq2.13f}).
}
\label{fig6}
\end{figure}

\begin{figure}
\includegraphics[keepaspectratio,scale=0.7]{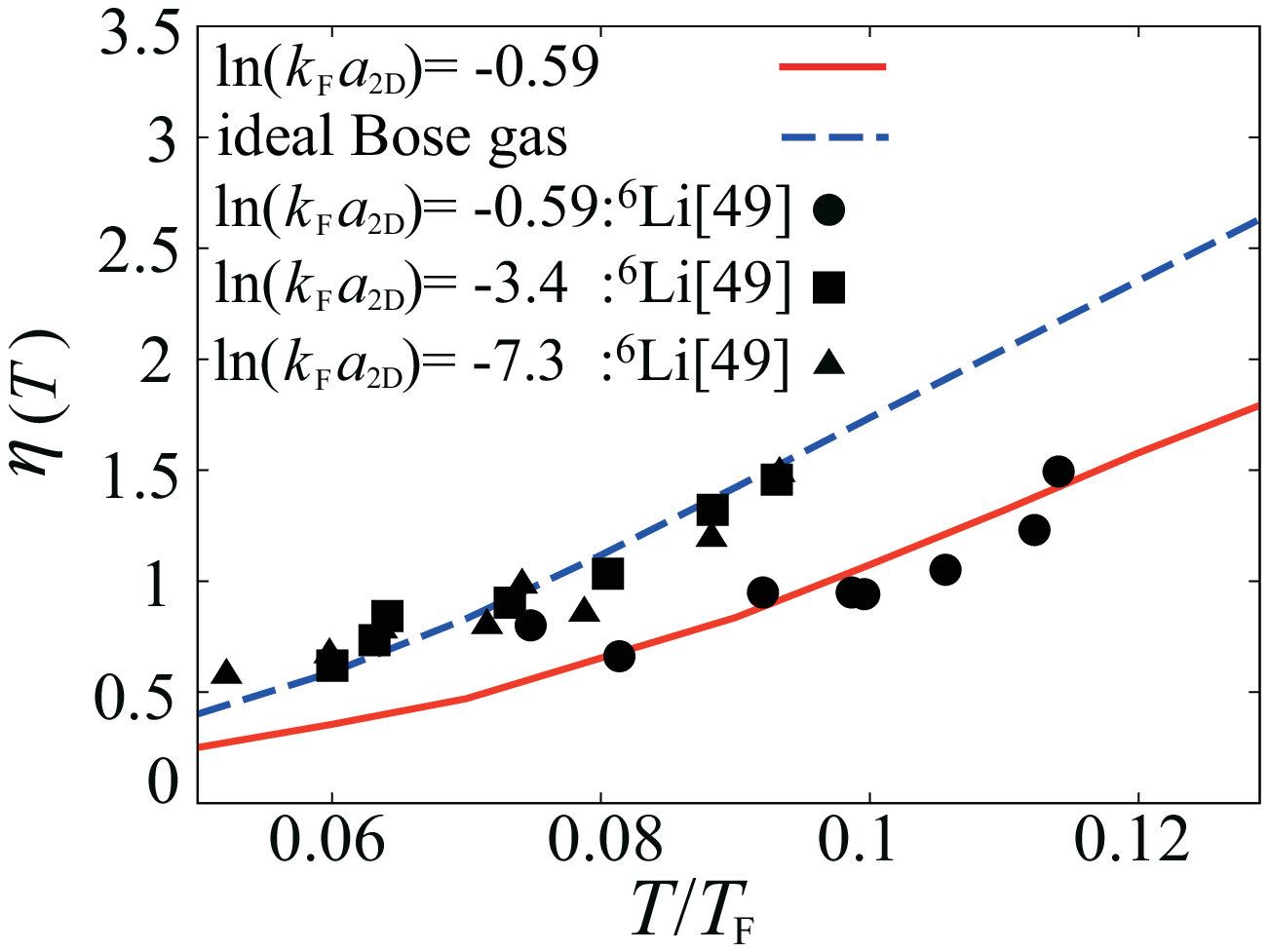}
\caption{(Color online) Calculated exponent $\eta$ in the assumed first-order correlation function $g_{1}(r)\propto 1/r^\eta$, which is determined by the fitting shown in Fig. \ref{fig6}. In this figure, ``ideal Bose gas" shows the result when the ordinary Bose distribution function is used in calculating $g_1(r)$. Since this result is almost the same as those when $\ln{(k_{\rm F}a_{\rm 2D})}=-3.4$ and $-7.3$, we do not explicitly show the latter two cases in this figure. We also show experimental results at $\ln{(k_{\rm F}a_{\rm 2D})}=-0.59$ (filled circles), $-3.4$ (filled squares), and $-7.3$ (filled triangles)\cite{Murthy1,notez}. These experimental data are also obtained from the spatial variation of $g_1(r)$ in the restricted spatial region, $5{\mu{\rm m}}\le r\le 25{\mu{\rm m}}$\cite{comment1}.
}
\label{fig7}
\end{figure}
\par
\section{First-order correlation function $g_1(r)$ of Cooper pair bosons}
\par
In Ref. \cite{Murthy1}, the exponent $\eta$ of the power-law decay in the first-order correlation function $g_1(r)\sim 1/r^\eta$ of Cooper-pairs is experimentally determined in the spatial region $5{\mu{\rm m}}\le r\le 25{\mu{\rm m}}$\cite{comment1}, in order to avoid effects of the edge of a gas cloud in a harmonic trap. When we also employ this prescription to theoretically evaluate $\eta$ in TMA, we obtain the fitting results at $\ln(k_{\rm F}a_{\rm 2D})=-0.59$, shown in Fig. \ref{fig6}. The resulting exponent $\eta$ is found to agree well with the recent experiment on a $^6$Li Fermi gas below $T_{\rm BKT}^{\rm exp}=0.11T_{\rm F}$\cite{Murthy1}, as shown in Fig. \ref{fig7}. In this figure, one also sees that our results agree with the experimental data in the stronger coupling cases, $\ln(k_{\rm F}a_{\rm 2D})=-3.4$ ($T_{\rm BKT}^{\rm exp}=0.1T_{\rm F}$) as well as $\ln(k_{\rm F}a_{\rm 2D})=-7.3$ ($T_{\rm BKT}^{\rm exp}=0.089T_{\rm F}$).
\par
As mentioned previously, the observed value $\eta\simeq 1.4$ at $T_{\rm BKT}^{\rm exp}$ is much larger than the prediction $\eta=0.25$ by the ordinary BKT theory\cite{Berezinskii1,Berezinskii2,Kosterlitz1,Kosterlitz2}. To understand why the present TMA approach reproduces the observed large value, it is useful to approximately evaluate the first-order correlation function $g_1(r)$ in Eq. (\ref{eq2.13}) by using Eq. (\ref{eq3.1a}), as
\begin{eqnarray}
g_1(r)
&\simeq& Z\sum_{\bm Q}
n_{\rm B}
\left(
{Q^2 \over 2M}-\mu_{\rm B}
\right)
e^{i{\bm Q}\cdot{\bm r}}
\nonumber
\\
&\simeq&
ZT
\int {d{\bm Q} \over 4\pi^2}
{1 \over {Q^2 \over 2M}-\mu_{\rm B}}
e^{i{\bm Q}\cdot{\bm r}}
\nonumber
\\
&=&
{MTZ \over \pi}K_0(\sqrt{2M|\mu_{\rm B}|}r),
\label{eq2.13f}
\end{eqnarray}
where $K_0(x)$ is the modified Bessel function. In the second line, we have assumed that the Bose distribution function is remarkably enhanced around ${\bm Q}={\bm 0}$. As shown in Fig. \ref{fig6}, Eq. (\ref{eq2.13f}) well describes the spatial variation of $g_1(r)$, especially in the ``experimental window," $5{\mu{\rm m}}\le r\le 25{\mu{\rm m}}$\cite{Murthy1,comment1}. Using the asymptotic form of the modified Bessel function $K_0(x\gg 1)$, we find 
\begin{equation}
g_1(r)\simeq
{MTZ \over \sqrt{2\pi}}
{1 \over \sqrt{\sqrt{2M|\mu_{\rm B}|}r}}
e^{-\sqrt{2M|\mu_{\rm B}|r}}.
\label{eq2.13g}
\end{equation}
Because $\mu_{\rm B}(T>0)$ in the two-dimensional case, Eq. (\ref{eq2.13g}) involves both the power-law decay ($1/r^{0.5}$) and the exponential decay ($e^{-\sqrt{2M|\mu_{\rm B}|}r}$). Thus, fitting this function to $g_1(r)\propto 1/r^\eta$ only in the ``experimental window," one obtains $\eta>0.5$ at $T_{\rm BKT}^{\rm exp}$, as seen in Fig. \ref{fig7}\cite{note2}.
\par
When $\ln(k_{\rm F}a_{\rm 2D})\lesssim -2$, the renormalization factor $Z$ almost equals unity, as shown in the inset in Fig. \ref{fig5}(b). As a result, the center-of-mass momentum distribution function $n_{\bm Q}$ of Cooper pairs is reduced to the Bose distribution function $n_{\rm B}(Q^2/(2M)-\mu_{\rm B})$, which no longer depends on the interaction strength. In Fig. \ref{fig7}, the dashed line corresponds to this situation, giving $\eta\simeq 1.4$ when $T\simeq 0.1T_{\rm F}$. This also naturally explains the reason why the experimental result on a $^6$Li Fermi gas\cite{Murthy1} ($\eta\simeq 1.4$ at $T_{\rm BKT}^{\rm exp}\simeq 0.1$) is almost independent of the interaction strength in the strong-coupling side.
\par
\section{Summary}
\par
To summarize, we have discussed low-temperature properties of a two-dimensional ultracold Fermi gas in the normal state. Including pairing fluctuations within a $T$-matrix approximation (TMA), we have calculated the center-of-mass momentum distribution $n_{\bm Q}$, as well as the first-order correlation function $g_1(r)$, of Cooper pairs in the strong-coupling side ($\ln(k_{\rm F}a_{\rm 2D})<0$).
\par
Recently, the remarkable amplification of the number $n_{{\bm Q}={\bm 0}}$ of Cooper pairs with zero center-of-mass-momentum was reported in a two-dimensional $^6$Li Fermi gas as a signature of the BKT phase transition\cite{Ries}. However, we clarified that this phenomenon can be quantitatively explained as a normal-state property of a strongly interacting two-dimensional Fermi gas, at least in the strong-coupling side. In this regime, the distribution function $n_{\bm Q}$ was shown to be well described by the Bose distribution function with the renormalization factor $Z$ describing pairing fluctuations.  Including the interaction dependence of $Z$, we showed that the temperature around which $n_{{\bm Q}={\bm 0}}$ starts to remarkably be enhanced increases with decreasing the interaction strength. This explains the observed behavior of the BKT phase transition temperature $T_{\rm BKT}^{\rm exp}$, which was determined as the temperature below which $n_{{\bm Q}={\bm 0}}$ remarkably increases. We briefly note that the interaction dependence of $T_{\rm BKT}^{\rm exp}$ is opposite to the prediction by the BKT theory that the BKT phase transition temperature should decrease with decreasing the interaction strength\cite{Berezinskii1,Berezinskii2,Kosterlitz1,Kosterlitz2}. In this regard, since we consider the normal state, our results do not contradict with this theoretical prediction.
\par
Within the same theoretical framework, we have also considered the origin of the observed large value of the exponent $\eta\simeq 1.4$\cite{Murthy1} in the first-order correlation function $g_1(r)\propto 1/r^\eta$ at $T_{\rm BKT}^{\rm exp}$. Although $g_1(r)$ in the normal state actually exhibits an exponential decay, when we fit the calculated first-order correlation function to $g_1(r)\propto 1/r^\eta$ in a restricted spatial region as done in the recent experiment\cite{Murthy1}, we found that the calculated $\eta$ well explains the observed large value $T_{\rm BKT}^{\rm exp}$. 
\par
Although our results do not immediately deny the recent realization of the BKT phase transition in a two-dimensional $^6$Li Fermi gas, we note that the some of the observed results do not agree with the ordinary BKT theory and the previous experiments on the BKT state in a two-dimensional Bose gas, as well as an exciton-polariton condensate. Since we clarified that these disagreements can be consistently resolved, if the system is still in the strongly interacting normal state, further experimental observations would be necessary, in order to unambiguously conclude that the BKT state is really realized in this system. 
\par
In this paper, we have only dealt with the normal state, to see to what extent the recent experiments\cite{Ries,Murthy1} can be explained as normal-state properties of a two-dimensional Fermi gas. To theoretically explore how to unambiguously detect the BKT transition within the current experimental technology in cold Fermi gas physics, extending the present TMA work to include the BKT transition is an interesting future problem. As mentioned at the end of Sec. II, inclusion of strong-coupling corrections to the particle-particle scattering matrix $\Gamma({\bm Q},i\nu_n)$ is also important to examine the weak-coupling side. Since the BKT phase transition is one of most exciting topics in cold Fermi gas physics, our results would be useful for the study of this two-dimensional Fermi superfluid.

\par
\begin{acknowledgments}
\par
We thank M. G. Ries and P. A. Murthy for useful comments, as well as sending their experimental data. We also thank R. Hanai, H. Tajima, T. Yamaguchi, and P. van Wyk for discussions. M. M. was supported by KLL PhD Program Research Grant, as well as Graduate School Doctoral Student Aid Program from Keio University. Y.O was supported by Grant-in-Aid for Scientific Research from MEXT and JSPS in Japan (No.25400418, No.15H00840). This work was supported by the KiPAS project in Keio university. 
\par
\end{acknowledgments}
\appendix
\section{Derivation of Eq. (\ref{eq3.1a})}
\par
In the strong-coupling regime where $\mu<0$ and the system is dominated by tightly bound molecules, expanding the pair correlation function $\Pi({\bm Q},i\nu_n)$ in Eq. (\ref{eq2.6}) in terms of ${\bm Q}$ and $\nu_n$ to $O(Q^2)$ and $O(\nu_n)$, respectively, we have
\begin{equation}
\Gamma(\bm{Q},\nu_{n})=\frac{8\pi|\mu|}{m}
\frac{1}{i\nu_{n}-\frac{Q^2}{2M}+\mu_{\rm B}}, 
\label{eqa.6}
\end{equation}
where $M=2m$, and 
\begin{equation}
\mu_{\rm B}=-{1 \over U}+\Pi(0,0)=-2|\mu| \ln \left(\frac{2|\mu|}{E_{\rm bind}} \right).
\label{eqa.7}
\end{equation}
In obtaining Eq. (\ref{eqa.6}), we have assumed $T\ll |\mu|$. At finite temperatures, since the superfluid phase transition does not occur in the two-dimensional case, the Thouless criterion in Eq. (\ref{eq2.7b}) is never satisfied, as $1-U\Pi(0,0)>0$. Thus, the Bose chemical potential $\mu_{\rm B}$ is always negative, as expected. 
\par
Substituting Eq. (\ref{eqa.6}) into Eq. (\ref{eq2.11}), and noting that Eq. (\ref{eqa.6}) becomes large around ${\bm Q}=\nu_n=0$, we obtain
\begin{eqnarray}
n_{\bm{Q}}&\simeq&
\left[
T\sum_{\bm{p},i\omega_{n}}
G_{0}(-\bm{p},-i\omega_{n})
G_{0}(\bm{p},i\omega_{n})
G(\bm{p},i\omega_{n})
\right]
T\sum_{i\nu_{n}}
\Gamma(\bm{Q},i\nu_{n}) 
\nonumber
\\
&=&-
\left[
T\sum_{\bm{p},i\omega_{n}}
G_{0}(-\bm{p},-i\omega_{n})
G_{0}(\bm{p},i\omega_{n})
G(\bm{p},i\omega_{n})
\right]
\frac{8\pi|\mu|}{m}
n_{\rm B}\left(\frac{Q^2}{2M}-\mu_{\rm B}\right).
\label{eqa.3}
\end{eqnarray}
Applying the same approximation to the TMA self-energy $\Sigma({\bm p},i\omega_n)$ in Eq. (\ref{eq2.4}), we obtain Eq. (\ref{eq2.13a}), leading to Eq. (\ref{eq2.13b}). Here, the pseudogap parameter $\Delta_{\rm PG}$ in Eq. (\ref{eq2.13c}) has the form,
\begin{equation}
\Delta_{\rm PG}=
\sqrt{
{8\pi|\mu| \over m}\sum_{\bm Q}
n_{\rm B}
\left(
{Q^2 \over 2M}-\mu_{\rm B}
\right)}.
\label{eqa.1b}
\end{equation}
Using Eq. (\ref{eq2.13b}), we carry out the $\omega_n$-summation in Eq. (\ref{eqa.3}). Then one obtains
\begin{equation}
n_{\bm Q}=
{
2|\mu|
(\sqrt{|\mu|^2+\Delta_{\rm PG}^2}-|\mu|)
\over 
\Delta_{\rm PG}^2
}
n_{\rm B}
\left(
{Q^2 \over 2M}-\mu_{\rm B}
\right).
\label{eqa2b}
\end{equation}
In the strong-coupling regime ($\mu<0$), the number $N_0$ of free Fermi atoms are almost absent at low temperatures ($T\ll|\mu|$), so that the number equation $N=N_0+\delta N$ is dominated by the fluctuation correction $\delta N=2\sum_{\bm Q}n_{\bm Q}$, as
\begin{eqnarray}
N\simeq \delta N
&=&
{
2|\mu|
[\sqrt{|\mu|^2+\Delta_{\rm PG}^2}-|\mu|]
\over 
\Delta_{\rm PG}^2
}
\sum_{\bm Q}
n_{\rm B}
\left(
{Q^2 \over 2M}-\mu_{\rm B}
\right)
\nonumber
\\
&=&
{m \over 4\pi}
\left[\sqrt{|\mu|^2+\Delta_{\rm PG}^2}-|\mu|\right],
\label{eqa2c}
\end{eqnarray}
where we have used Eq. (\ref{eqa2b}) in obtaining the last expression. Substituting $N=k_{\rm F}^2/(2\pi)$ into Eq. (\ref{eqa2c}), one finds $\Delta_{\rm PG}=2\sqrt{\varepsilon_{\rm F}(\varepsilon_{\rm F}+|\mu|)}$. The substitution of this result into Eq. (\ref{eqa2b}) gives Eq. (\ref{eq3.1a}).
\par
In the low temperature limit $T\to 0$, the right hand side of the first line in Eq. (\ref{eqa2c}) is finite only when $\mu_{\rm B}\to 0$, so that one finds $\mu(T\to 0)=-E^{\rm 2D}_{\rm bind}/2$.
\par
\par


\begin{thebibliography}{9}
\bibitem{notez} In Refs. \cite{Ries,Murthy1}, the interaction strength slightly varies with decreasing the temperature. In the legend in Fig. \ref{fig7}, the interaction strengths for experimental data are the values at $T_{\rm BKT}^{\rm exp}$.
\bibitem{note2} At low temperatures, since $\mu_{\rm B}$ becomes very small (see Fig. \ref{fig5}(a)), one finds that $\sqrt{2M|\mu_{\rm B}|}r<1$ in the experimental window, $5{\mu{\rm m}}\le r\le 25{\mu{\rm m}}$. In this case, Eq. (\ref{eq2.13g}) is no long valid, so that the exponent $\eta$ may be smaller than 0.5, as shown in Fig. \ref{fig7}.
\end{thebibliography}
\end{document}